\documentclass[
 reprint,
]{revtex4-2}

\usepackage{graphicx}
\usepackage{dcolumn,footnote}
\usepackage{comment,amssymb,amsmath}       %
\usepackage{graphicx}                     %
\usepackage{longtable,comment}            %
\usepackage[colorlinks = true,            %
            linkcolor = blue,             %
            urlcolor  = blue,             %
            citecolor = blue,             %
            anchorcolor = blue]           %
            {hyperref}                    %

\begin{document}

\title{
Invariant density statistical quantifiers and a temperature for the logistic map
}

\author{Samuel S. Santos}
 \email{202410487@uesb.edu.br}
\affiliation{Departamento de Ciências Exatas e Naturais, Universidade Estadual do Sudoeste da Bahia,
			    BR 415, Itapetinga - BA, 45700-000, Brazil}

\author{Guilherme V. B. Junior}%
 \email{202011272@uesb.edu.br}
\affiliation{Departamento de Ciências Exatas e Naturais, Universidade Estadual do Sudoeste da Bahia,
			    BR 415, Itapetinga - BA, 45700-000, Brazil}

\author{Ignacio S. Gomez}%
 \email{ignacio.gomez@uesb.edu.br}
\affiliation{Departamento de Ciências Exatas e Naturais, Universidade Estadual do Sudoeste da Bahia,
			    BR 415, Itapetinga - BA, 45700-000, Brazil}

\author{Zolacir T. Oliveira Junior}

\email{ztoliveira@uesc.br}

\affiliation{Departamento de Ciencias Exatas e Tecnológicas, Universidade Estadual de Santa Cruz, Ilhéus, 45662 000, Bahia, Brazil}

\author{Ronaldo Thibes}

\email{thibes@uesb.edu.br}

\affiliation{Departamento de Ciências Exatas e Naturais, Universidade Estadual do Sudoeste da Bahia,
			    BR 415, Itapetinga - BA, 45700-000, Brazil}
\begin{abstract}
In this work, we study the dynamics of the logistic map 
based on a probabilistic characterization in terms of the invariant density.
We analyze the relevant regimes of the dynamics (regular, oscillatory, onset chaotic and fully chaotic) in terms of the Fisher information and the Crámer-Rao (CR) complexity. We have found that these informational quantifiers allow to distinguish the dynamical regions of the map, by maximizing the Fisher information in the regular behavior and with the CR complexity exhibiting variations and a maximum near to the Pameau-Maneville scenario. 
Fisher information as a function of time is examined in the light of Frieden's informational interpretation of the Second Law of Thermodynamics. 
We apply the Equipartition Theorem to propose a definition of temperature for the logistic map, providing a macroscopic signature of the dynamics.      
\end{abstract}

\maketitle

\section{Introduction}
Discrete maps of the form $x_{i+1}=f(x_i)$ constitute a powerful tool in the study of dynamical systems. 
Among those, the logistic map stands out as one of the most interesting ones, possessing several applications in physics, biology, chemistry, engineering, econophysics, mathematics and science in general \cite{May-1976, Ausloos-Dirickx-2006,Strogatz,Lloyd}. 
The iteration of its simple quadratic formula reveals an unexpected richness allowing to describe distinct dynamical regimes going from regular to fully chaotic ones, passing in between through intermittency, aperiodicity and regular-chaotic mixed behaviors,  
manifesting its ability for testing chaos, ergodicity, onset of chaos, complexity and serving as building block for more complex model constructions regarding population growth, epidemic control, quantum systems and many others \cite{Tirnakli-2009,Tsallis-book,Pessoa-2011,Csordas-1993}. 
Complementarily, a probabilistic description for discrete systems is given by the theory of invariant measures and Markov operators \cite{Lasota-1985,Beck-Schogl,Walters}. In that approach,
the dynamical system features 
are studied from the probability densities evolution characterizing the knowledge of the system at a given instant. The evolution of any density is obtained from the iterations of the Frobenius-Perron operator over an arbitrary initial density. Of particular importance is the existence of invariant densities of the Frobenius-Perron operator since they represent stationary states modeling the system in the asymptotic limit of large times.

On the other hand, statistical quantifiers have been shown to be important theoretical tools for characterizing several phenomena in physics and related areas.
Notably, the Fisher information (FI) concept \cite{Fisher-1922, Frieden} 
can be employed for exploring several phenomena like statistical ensembles, environmental systems, quantum states, information in inhomogeneous systems, etc
\cite{Plastino, Mayer,Gomez-2019,Costa-2020}. FI arose from the problem of parameter estimation in the context of statistics, and since then it has been employed for characterizing pattern genetics in biology, dynamical phase transitions in environmental engineering, chemistry, economy, quantum mechanics, deduction of motion equations, by making use of its sensibility against local variations of the probability distribution. 
For a given one-dimensional continuous probability density function $p(s)$, the corresponding Fisher information $I$ can be defined as (Eq. (1.2) of \cite{Frieden})
\begin{equation}\label{FI-continuous}
I = \int dx \, {p'(x)}^2/p(x)\,.
\end{equation}
Intuitively, FI can distinguish situations with behaviors being radically different. For instance, 
a regular behavior characterized by a piked distribution gives a high value for $I$, while a fully chaotic dynamics represented by a uniform distribution minimizes FI.
The Crámer-Rao (CR) complexity plays a complementary role by distinguishing behaviors where FI alone is not sufficient. When we have different dynamical regimes with similar values for the Fisher information, we can calculate the CR complexity for distinguishing between them.
Additionally, FI has been studied to give an alternative formulation of the Second Law of Thermodynamics. 
Illustrating with several examples and theoretical 
arguments, in \cite{Frieden}, Frieden has suggested that FI can provide a interpretation of the arrow of time, by postulating that Second Law of Thermodynamics is equivalent to the condition $dI/dt\leq 0$. Physically, this inequality means that if the FI level of a system is observed to be decreasing, that is $dI<0$,
then its history is necessarily advancing, which is the
main content of the “$I$-Theorem” (Section 1.8.2 of \cite{Frieden}).

The main goal of the present work is to characterize some relevant regimes of the logistic map dynamics from a probabilistic description in terms of the invariant density, from which we calculate FI and associated CR complexity for analyzing dynamical transitions. In addition, we propose a notion of temperature based on the Equipartition Theorem, which allows to provide a macroscopic signature of the map and an analysis of the Frieden's informational interpretation of the Second Law. The work is organized as follows. In Section II we give the preliminaries. Section III is devoted to the calculation of the Fisher information and the Crámer-Rao complexity, along with our definition of temperature for the logistic map (called map temperature MP). Here we perform numerical simulations of the statistical quantifiers and the MP in function of the parameter of the map and of the time step. Finally, in Section IV, we outline some conclusions and perspectives. 

\section{Preliminaries}

We begin by presenting the minimal concepts and methods to be employed in the subsequent Sections of the paper. 

\subsection{Probability density functions, discrete maps and invariant density}

A probability density function defined over a topological
space $\Gamma$ (typically, a subset of $\mathbb{R}^m$) is any nonnegative function $p:\Gamma\rightarrow\mathbb{R}_+$ such that $\int_\Gamma p(x)dx=1$ with $\Gamma$ called the \emph{space of events}. In an experiment, $\Gamma$ represents the set of all possible outcomes. If these ones are $W$ possible results, then one has a discrete probability distribution expressed by a probability vector $(p_1,p_2,\ldots,p_W)$, being $p_i$ the probability that the $i$-th result occurs and the normalization condition reads $\sum_{i=1}^Wp_i=1$.

Given $\Gamma$ and a continuous function $f:\Gamma\rightarrow \Gamma$, we say that the sequence 
$\{x_n\}_{n\in \mathbb{N}_0}\subseteq \Gamma$
such that
\begin{equation}\label{map-def}
    x_{n+1}=f(x_n) \quad \forall n\in \mathbb{N}_0
\end{equation}
defines a \emph{discrete map}. Given $x_0\in \Gamma$ as the starting point, the sets $\{x_n: n\in \mathbb{N}_0\}$ are called \emph{orbits}, with $x_{\infty}=\lim_{n\rightarrow \infty} x_n$ standing for the limit of an orbit. From the physical viewpoint, a
discrete map models a system whose dynamics is
given by iterating \eqref{map-def}, where each iteration corresponds to a time step. If the system is initially in a state $x_0$, then $x_n$ represents the state after $n$ time steps. When there exists an element 
$x^*$ such that $f(x^*)=x^*$, we say that $x^*$ is a \emph{fixed point} of $f$, which can be physically interpreted as a stationary state of the system. The dynamics of a discrete map can be characterized in terms of probability density functions as follows. The Frobenius–Perron operator $P:\mathbb{L}^1(\Gamma)\rightarrow\mathbb{L}^1(\Gamma)$ associated with the map \eqref{map-def} is given by \cite{Lasota-1985,Beck-Schogl,Walters}
\begin{equation}\label{FPoperator}
    \int_A P\phi(x)dx=\int_{f^{-1}(A)}\phi(x)dx
\end{equation}
for all $\phi\in \mathbb{L}^1(\Gamma)$ and $A\subseteq \Gamma$ with $f^{-1}(A)$ the preimage of $A$ and 
$\mathbb{L}^1(\Gamma)$ the set $\{g:\Gamma\rightarrow \mathbb{R} \ \textrm{with} \ \int_\Gamma  | g(x)|dx<\infty\}$. Any non-negative normalized function $\rho\in \mathbb{L}^1(\Gamma)$ 
with
\begin{equation}\label{invariant-density}
    P\rho(x) = \rho(x) \quad \forall x\in \Gamma
\end{equation}
is called the \emph{invariant density} of the map, constituting a invariant measure \cite{Walters} that physically represents the state of the system in the asymptotic limit of large iterations \cite{Lasota-1985}. 
Numerically, the invariant density $\rho(x)$ can be constructed by
dividing $\Gamma$ into $W$ intervals and defining  
\begin{equation}\label{invariant-density}
    \rho_i = 
    \frac{\# \{x_\tau \in [\frac{i-1}{W},\frac{i}{W}) \ | \ \tau=0,1,\ldots,N\}}{N}
\end{equation}
as the fraction of points $x_\tau$ in the interval $[\frac{i-1}{W},\frac{i}{W})$ generated by iterations of $f(x)$ from an initial point $x_0\in \Gamma$. Thus, when $N\gg1$ and $N\gg W$ the invariant density $\rho(x)$ results well approximated numerically by 
\begin{equation}\label{numerical-invariant-density}
    \rho(x) = \sum_{i=1}^W \rho_i 1_{[\frac{i-1}{W},\frac{i}{W})}(x) 
\end{equation}
with $1_{[\frac{i-1}{W},\frac{i}{W})}(x)$ standing for the characteristic function of the interval $[\frac{i-1}{W},\frac{i}{W})$. It is worth to be noted that in the limit $N\rightarrow \infty$ the invariant density $\rho(x)$ does not depend on the starting point $x_0$. Recently, numerical invariant density has been employed to define a statistical distance in discrete maps \cite{Gomez-2017}, which resembles the Wooter's distance \cite{Wooters} for distinguish quantum states. 

\subsection{Logistic map}

The \emph{logistic map} is given by \cite{May-1976}
\begin{equation}\label{logistic-map}
    x_{n+1}=\mu x_n(1-x_n)  \quad , \quad n\in \mathbb{N}_0
\end{equation}
where $\mu\in(0,4]$ is the external parameter and $x_0\in\Gamma=[0,1]$ the initial condition. The case $\mu=0$ is not of interest since we have $x_{n+1}=0$ for all $n\in \mathbb{N}_0$. 
We review some relevant regimes that we will employ for characterizing them from the formalism presented. When $0<\mu\leq 3$ all the orbits are convergent and we have that 
$x_{\infty}=0$ for $0<\mu<1$ and $x_{\infty}=\frac{\mu-1}{\mu}$ for $1\leq \mu\leq 3$ with different rates of convergence. For $0<\mu< 3$ the orbits rapidly approach to the asymptotic limit with a linear convergence for $2\leq \mu <3$ and less than linear for $\mu=3$. When $3\leq \mu\leq 3.44949$ the orbits oscillate between two values and with $3.44949<\mu\leq 3.56995$ almost all the orbits oscillate between four values. The value $\mu=3.56995$ represents the beginning of the chaotic behavior (onset of chaos) where orbit oscillations of finite period are not observed. The region $3.56995\leq \mu \leq 3.82843$ represents the called \emph{Pameau-Maneville scenario} \cite{Pomeau-1980}, with the orbits manifesting a periodic laminar phase interrupted by bursts of aperiodic behavior. For $\mu=4$ the fully chaotic behavior emerges with all the orbits dense in $[0,1]$ provided with a mixing dynamics. Interestingly, when $\mu=2$ and $\mu=4$ there exist exact solutions for the orbits, given by
\begin{equation}\label{orbit-formula}
    x_n=
    \begin{cases}
      \frac{1-\exp{(2^n\log(1-x_0))}}{2}& \textrm{for}\quad \mu=2 \\
      \frac{1-\cos{(2^n\arccos(1-x_0))}}{2}&  \textrm{for} \quad \mu=4.
    \end{cases}
\end{equation}
The invariant density has a closed formula for $\mu=4$. Using \eqref{FPoperator}, in this case we have for all $\phi \in \mathbb{L}^1([0,1])$ 
\begin{equation}\label{density-chaotic-1}
    P\phi(x) = \frac{\phi \Bigg(\frac{1}{2}+\frac{1}{2}\sqrt{1-x} \Bigg) + \phi \Bigg(\frac{1}{2}-\frac{1}{2}\sqrt{1-x} \Bigg)}{4\sqrt{1-x}},
\end{equation}
so in the limit of infinite successive iterations $N\rightarrow\infty$ we obtain the invariant density $\rho(x)$
\begin{equation}\label{density-chaotic-1}
    \rho(x) = \lim_{N\rightarrow\infty} P^N\phi(x) = \frac{1}{\pi\sqrt{x(1-x)}}.
\end{equation}
It should to be noted that, except for particular cases, the invariant density has no analytical closed formula so in general we have to compute it only numerically. 

\subsection{Fisher information and Crámer-Rao complexity}

The discrete version of the Fisher information (FI) \eqref{FI-continuous} is given in terms of the expression 
\begin{equation}\label{Fisher-information}
    I[p]=4\sum_{i=0}^{W-1}
    \Big(\sqrt{p_{i+1}} - \sqrt{p_{i}}\Big)^2
\end{equation}
for a given discrete probability density $\{p_i\}$, $p_i\geq 0$, normalized as $\sum p_i=1$.
Associated to (\ref{Fisher-information}), we define the Crámer-Rao complexity
\begin{equation}\label{CR}
    C[p] = I[p] \times \sigma^2 
\end{equation}
with $\sigma^2$ standing for the variance of the probability density $\{p_i\}$, that is 
\begin{equation}
    \sigma^2 =  
    \sum_{i=1}^W z_i^2p_i - \Big(
    \sum_{i=1}^W z_ip_i \Big)^2 
\end{equation}
and $z_i\in [\frac{i-1}{W},\frac{i}{W})$ with $p = (p_1,p_2,\ldots,p_W)$ a discrete probability density. 
As mentioned in the Introduction, the  Crámer-Rao complexity
is an important statistical quantifier complementary to the Fisher information. 
The Crámer-Rao bound theorem \cite{Frieden} states that the CR complexity is low bounded by 1, i.e. $C[p]\geq 1$, being maximized for Gaussian probability densities.  Here, we only focus on FI and the corresponding CR complexity, being other information measures and their associated complexities \cite{LMC} out of the scope of the present work. 


\subsection{Equipartition Theorem}

The
Equipartition Theorem (ET) establishes a relation between a system's temperature and its constituent particles total kinetic energy average, expressed in the general form by
\begin{equation}\label{EQT}
    \Bigg\langle q_m \frac{\partial H}{\partial q_n}\Bigg\rangle = \delta_{mn}k_B T
\end{equation}
where the brackets mean an ensemble average obtained from the Liouville equilibrium density in the asymptotic limit $t\rightarrow \infty$, $H$ is the Hamiltonian of the system, $k_B$ the Boltzmann constant, $T$ its temperature and $q_n$ denotes the $n$-th coordinate of the system in phase space. For the special case of a system composed by non-interacting particles of mass $m$, the average energy per particle $E_{\textrm{particle}}=\langle H_{\textrm{particle}}\rangle$ is given by the Hamiltonian $H_{particle}=(p_x^2+p_y^2+p_z^2)/2m$, so applying  
\eqref{EQT} for $q_i=p_i$ ($i=x, y, z$) it follows 
\begin{equation}\label{EQT-gas}
E_{\textrm{particle}}
    = 
    \frac{m}{2}\langle v^2\rangle=
    \frac{1}{2}\sum_{i=x,y,z}\Bigg\langle  p_i\frac{\partial H}{\partial p_i}\Bigg\rangle
    =
    \frac{3}{2}k_B T,
\end{equation}
from which we can see that the temperature of the system is proportional to the average energy per particle. 
It is worth noting that some hypotheses are needed for the validity of the ET. Typically, it holds for \emph{ergodic} systems in thermal equilibrium with a density distribution in phase space equiprobable for all the states. For our proposal, it is convenient to make \eqref{EQT-gas} dimensionless, so considering the one-dimensional case, from \eqref{EQT-gas} we can deduce the dimensionless temperature $T/T_0$ as 
\begin{equation}\label{EQT-dimensionless}
    \frac{T}{T_0}= \frac{\langle v^2\rangle}{v_0^2}\frac{m v_0^2}{k_BT_0},
\end{equation}
where $T_0$ and $v_0$ denote two arbitrary characteristics temperature and velocity for the system. 

\subsection{Frieden's informational interpretation of the Second Law of thermodynamics}
In order to be compatible the arrow of time with the Fisher information, in Eq. (1.30) of \cite{Frieden} it is has been postulated 
\begin{eqnarray}\label{FI-arrow}
    \frac{d I}{dt}\leq 0,
\end{eqnarray}
which 
is equivalent to the H-Theorem $\frac{dH}{dt} \geq0$ for linear Fokker-Planck equations. 
Physically, equation \eqref{FI-arrow} says that if the Fisher information level of a system is observed to be decreasing, $dI < 0$, then its history is necessarily advancing, $dt > 0$ (I-Theorem, Section 1.8.2 of \cite{Frieden}). In what follows, equation \eqref{FI-arrow} will be referred as Frieden's informational interpretation of the Second Law. 

\section{Statistical quantifiers and temperature for logistic map}
In this section,
we calculate the FI and its corresponding CR complexity for a set of values of the parameter $\mu$ of the logistic map.  
After that, we analyze the temporal evolution of the FI from the Frieden's interpretation of Second Law of Thermodynamics for some representative values of $\mu$. Finally, by employing the ET, an associated temperature for the logistic map is obtained along with the FI and the CR complexity as a function of the temperature.  

\subsection{FI and CR complexity for the logistic map as functions of the map parameter}
For the numerical calculations of FI and CR complexity, we employed $N=10^6$ steps and $W=10^4$ bins, thus guaranteeing $N\gg W$. By increasing $N$,
the results remained the same due to the convergence of the invariant density.
In Figs. 1 and 2, we illustrate the FI and the CR complexity as functions of the parameter $\mu$.
\begin{figure}
    \centering
\includegraphics[width=1\linewidth]{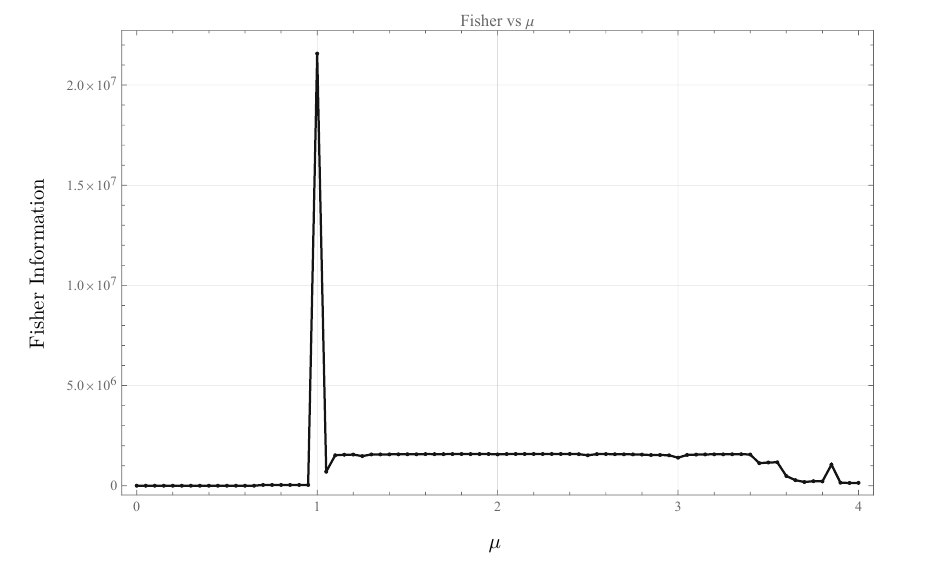}
    \caption{Fisher information \eqref{Fisher-information} of the invariant density \eqref{numerical-invariant-density} of the logistic map as a function of the parameter $\mu$ for $\mu=0.05, 0.1,\ldots, 3.9, 3.95, 4$ with $N=10^6$ number of steps and $W=10^4$ bins.}
    \label{fisherxmu}
\includegraphics[width=1\linewidth]{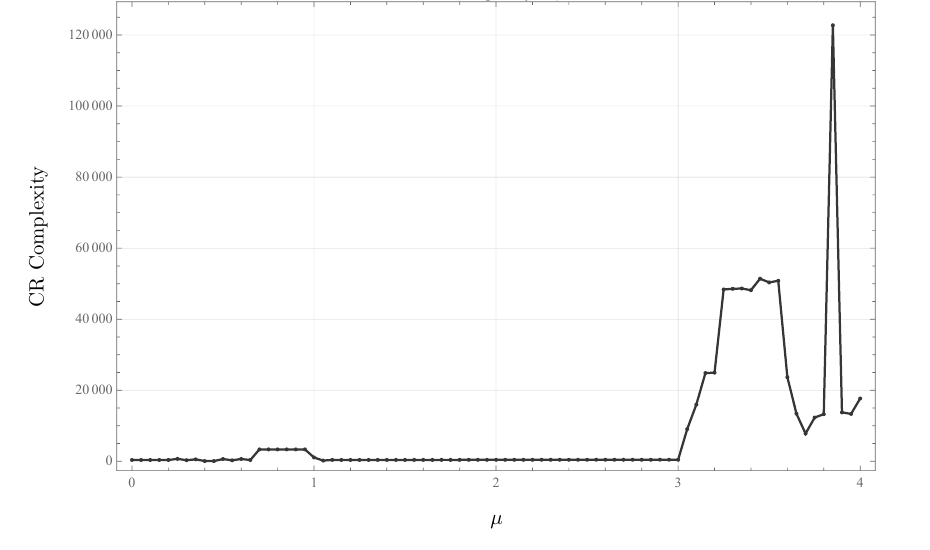}
    \caption{Crámer-Rao complexity \eqref{CR} of the invariant density \eqref{numerical-invariant-density} of the logistic map as a function of the parameter $\mu$ for $\mu=0.05, 0.1,\ldots, 3.9, 3.95, 4$ with $N=10^6$ number of steps and $W=10^4$ bins.}
    \label{complexityxmu}
\end{figure}
From Fig. \ref{fisherxmu}, we see that FI presents an absolute maximum value for $\mu=1$, which corresponds to regular behavior with orbits converging to $x_\infty=0$. For $0\leq \mu <1$ and $1\leq \mu <3.4$ the FI behaves approximately constant, from which it is not possible to distinguish the dynamical transitions for these values. From $\mu=3.4$ the FI decreases up to the region $3.8< \mu<3.9$ where a local maximum occurs. FI in the fully chaotic regime $\mu=4$ has the same value as for $3.7\leq \mu\leq 3.8$. 
By contrast, from Fig. \ref{complexityxmu} we see that the CR complexity presents a variation in the region $3\leq \mu\leq 4$, which includes the onset of chaos zone from $\mu=3.56995$. A peak is observed between $\mu=3.8$ and $\mu=4$, thus manifesting a detection of the periodic laminar phase with aperiodic behavior characteristic of the \emph{Pameau-Maneville scenario}, that is a maximal complex behavior.
In Fig. \ref{complexityxfisher}, we see how the CR complexity takes high values while the FI is vanishingly small and vice-versa. \begin{figure}
    \centering
\includegraphics[width=1\linewidth]{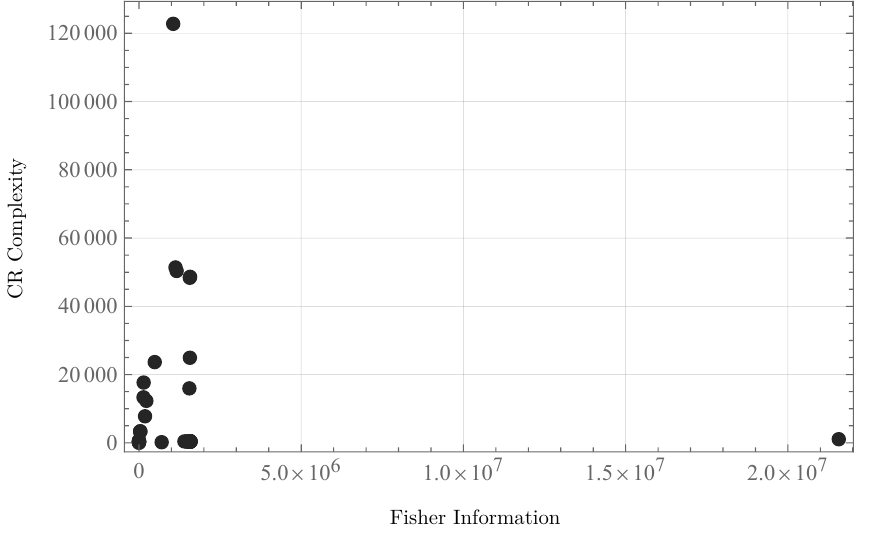}
\caption{CR complexity \eqref{CR} as a function of the Fisher information \eqref{Fisher-information} for the invariant density \eqref{numerical-invariant-density} of the logistic map with $\mu=0.05, 0.1,\ldots, 3.9, 3.95, 4$ with $N=10^6$ number of steps and $W=10^4$ bins.}
\label{complexityxfisher}
\end{figure}

\subsection{Logistic map FI and CR time evolution}
With the aim of testing Frieden's informational interpretation of the Second Law \eqref{FI-arrow}, now we analyze the FI and the CR complexity as functions of the time steps. The results are exhibited in Figs. 4 and 5
for regular $\mu=1$, 
onset of chaos $\mu=3.56995$ and fully chaotic $\mu=4$ regimes.  For both FI and CR complexity, we can see that after a short transient, a monotonic behavior is manifested, except for the regular case $\mu=1$, and with appreciable fluctuations for the fully chaotic regime $\mu=4$. When $\mu=1$, we observe that FI is decreasing only by short periods of time, after which it grows abruptly. Hence, the behavior of the FI in the regular regime $\mu=1$ represents a violation of the Frieden's interpretation of the Second Law $dI/dt\leq 0$ \eqref{FI-arrow}. On the other hand, the CR complexity behaves monotonically decreasing for $\mu=1$, as we can see from upper panel of Fig. 5.

\begin{figure}
\begin{minipage}[h]{1\linewidth}
\includegraphics[width=\linewidth ]{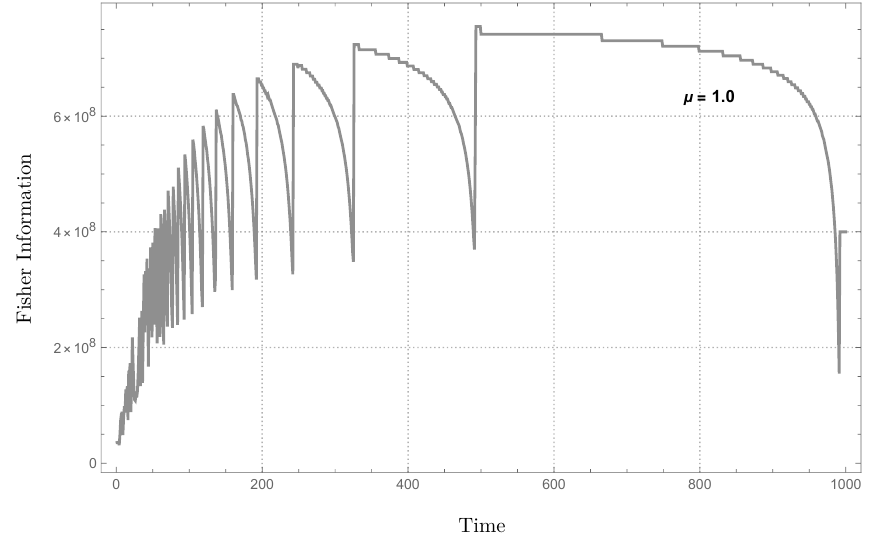}
\end{minipage}
\begin{minipage}[h]{1\linewidth}
\includegraphics[width=\linewidth]{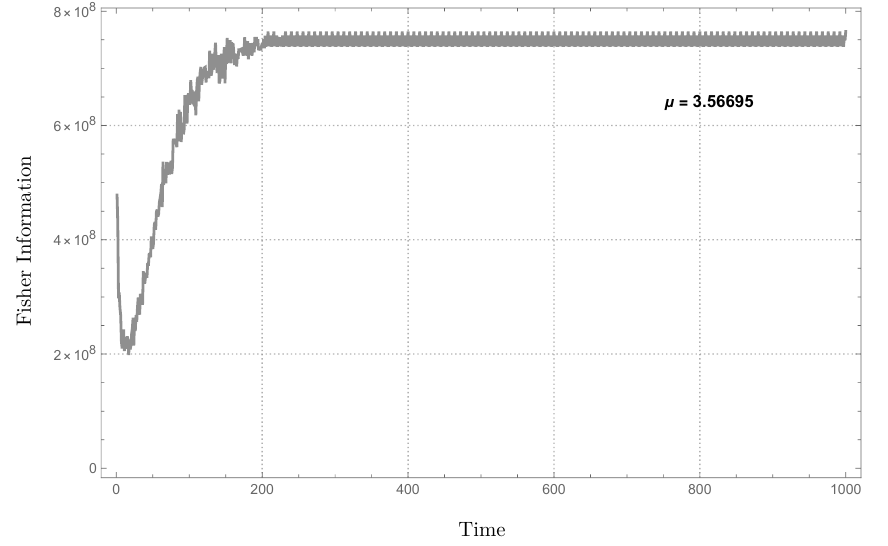}
\end{minipage}
\begin{minipage}[h]{1\linewidth}
\includegraphics[width=\linewidth]{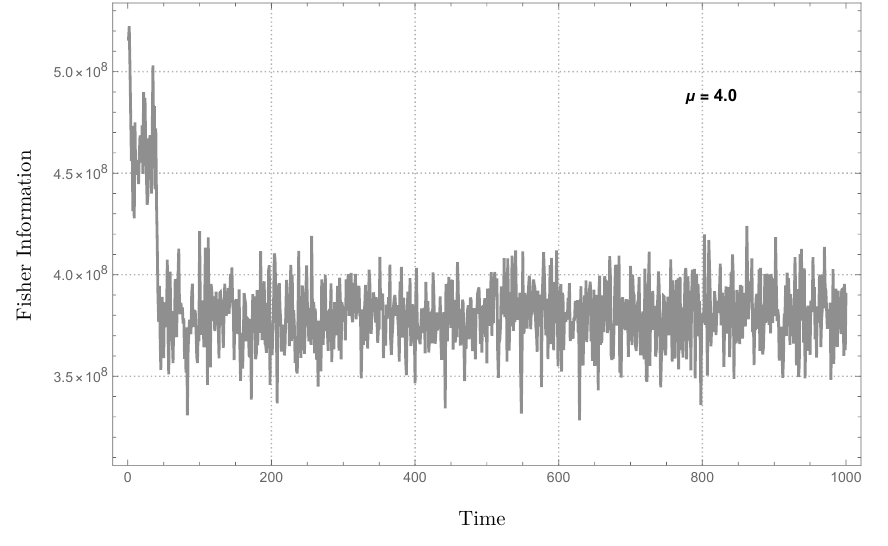}
\end{minipage}
\caption{Fisher information versus time step of the logistic map for the parameter values $\mu=1$ (upper), $\mu=3.56695$ (center) and $\mu=4$ (bottom) with $N=10^6$ number of steps and $W=10^4$ bins.}
\end{figure}

\begin{figure}
\begin{minipage}[h]{1\linewidth}
\includegraphics[width=\linewidth ]{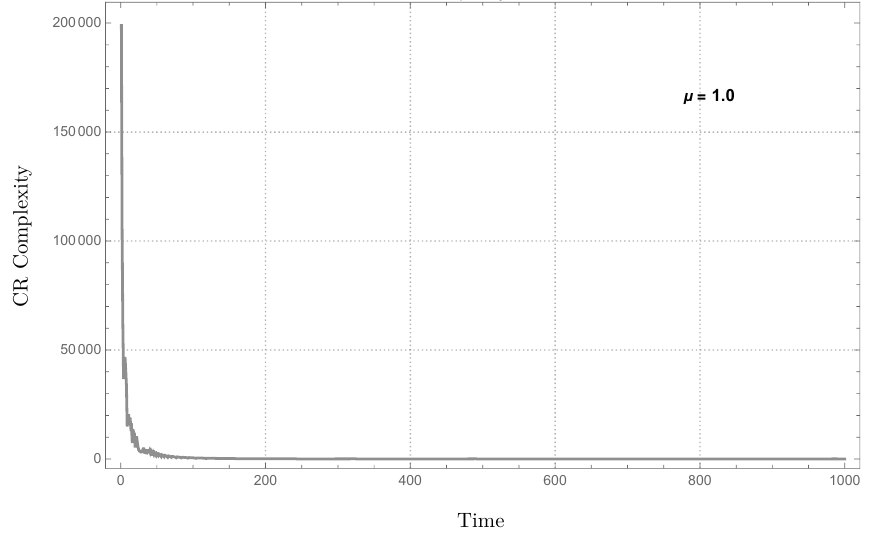}
\end{minipage}
\begin{minipage}[h]{1\linewidth}
\includegraphics[width=\linewidth]{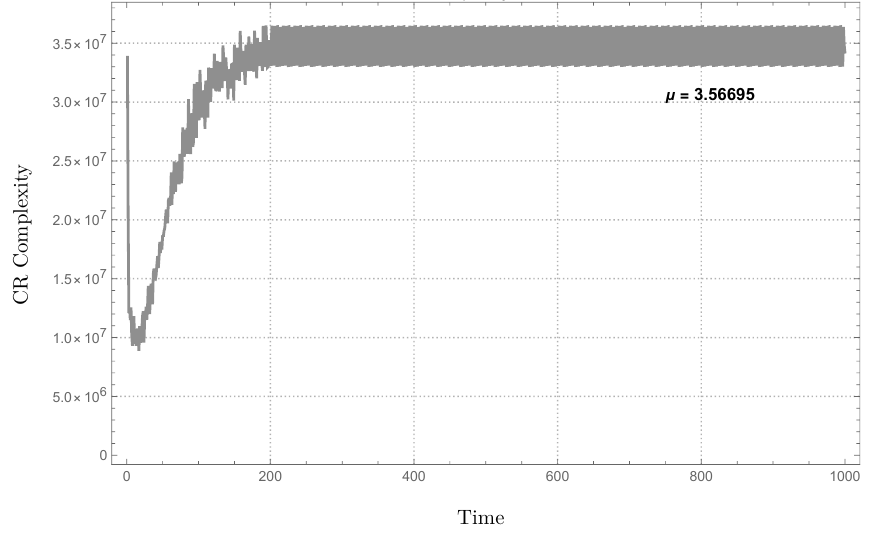}
\end{minipage}
\begin{minipage}[h]{1\linewidth}
\includegraphics[width=\linewidth]{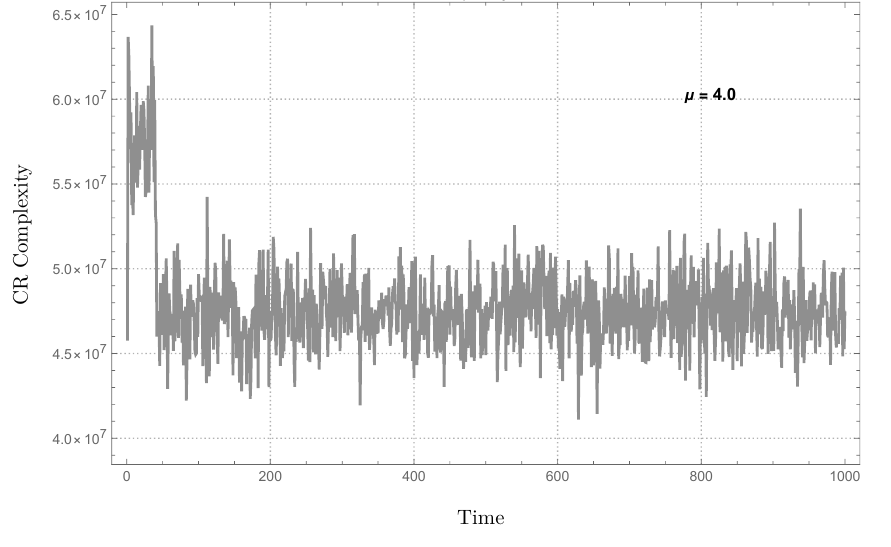}
\end{minipage}
\caption{Crámer-Rao complexity versus time step of the logistic map for the parameter values $\mu=1$ (upper), $\mu=3.56695$ (center) and $\mu=4$ (bottom) with $N=10^6$ number of steps and $W=10^4$ bins.}
\end{figure}

\subsection{A temperature associated to the logistic map}
Considering that $v_i=x_{i+i}-x_i$ represents a dimensionless average velocity for the logistic map between the $i$-th and $(i+1)$-th steps, we apply the relationship \eqref{EQT-dimensionless} with $mv_0^2/(k_BT_0)=1$ and propose the \emph{map temperature} (MP) at $N$-th step, denoted by $T(\mu,N,M)$ for the logistic map as follows:
\begin{equation}\label{MP}
    T(\mu,N,M)= \frac{1}{M}\sum_{j=1}^M (x_{N+1}^{(j)}-x_N^{(j)})^2     \quad \forall \  N=0,1,\ldots,
\end{equation}
where $\{x_k^{(j)}\}$ is the orbit starting at $x_0^{(j)}$ with $j=1,\ldots,M$. 
The physical meaning of the expression \eqref{MP} is that $T(\mu,N,M)$ characterizes the transition dynamics of the logistic map at the $N$-th step, averaged over the $M$ initial conditions belonging to the space state $[0,1]$.

\begin{figure}
    \centering
    \includegraphics[width=1\linewidth]{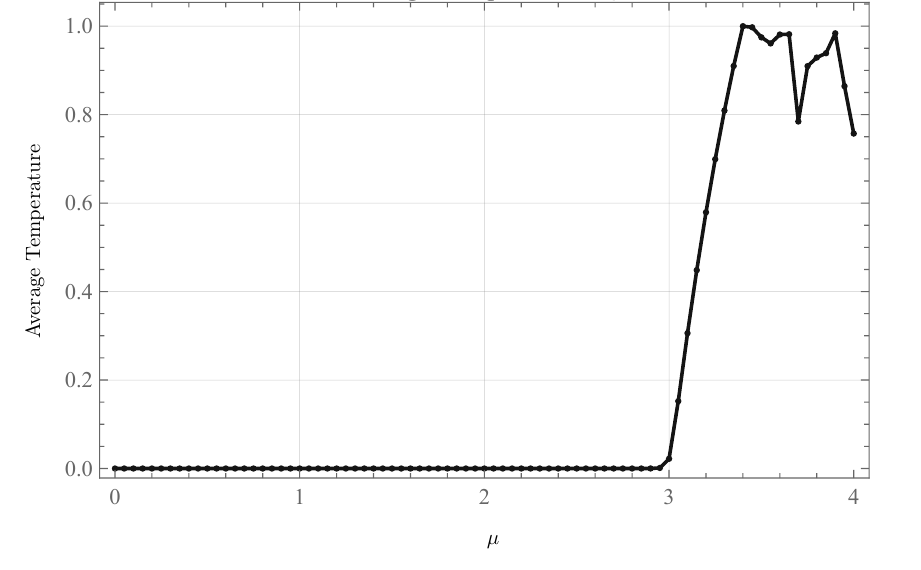}
    \caption{The map temperature \eqref{MP} for the values of the map parameter $\mu=0.05, 0.1, \ldots, 3.9,  3.95, 4$ with $N=10^3$ number of steps and $M=10^5$ initial conditions is illustrated.}
    \label{averagetemperature}
\end{figure}

We calculate MP \eqref{MP} for the same range of values of $\mu=0.05, 0.1, \ldots, 3.95, 4$ as performed in Section III.A, with $M=10^5$ initial conditions and $N=10^3$. It is worth to mention that for $M>10^5$ the MP remained with the same behavior. 
When calculating MP from definition \eqref{MP}, we observed the existence of a short time transient along with the presence of small variations of the $T(\mu,N,M)$ for some values of $\mu$. In order to define a temperature for a large number of steps we have taken the average of \eqref{MP} neglecting the transient. 
The dependence of the MP in function of $\mu$ is shown in Fig. \ref{averagetemperature}. For a better visualization, the MP was normalized according to $T(\mu,N,M)/\max\{T(\mu,N,M)\}$ with the the maximum taken for $N\geq N_0$, being $N\leq N_0$ the transient region.
We also can analyze the behavior of the map temperature $T(\mu,N,M)$ as a function of the time step, which is illustrated in Fig. 7. It is observed that an associated temperature is well defined for the regimes with less complexity, that is, the regular $\mu=1$ and the fully chaotic $\mu=4$ ones. By contrast, when the regime presents more complexity, $\mu=3.56695$ in our case, the temperature exhibits non trivial fluctuations that prevents to define it. In turn, this can be understood from the manifestation of the complex behavior of the onset of chaos.

\begin{figure}
\begin{minipage}[h]{1\linewidth}
\includegraphics[width=\linewidth ]{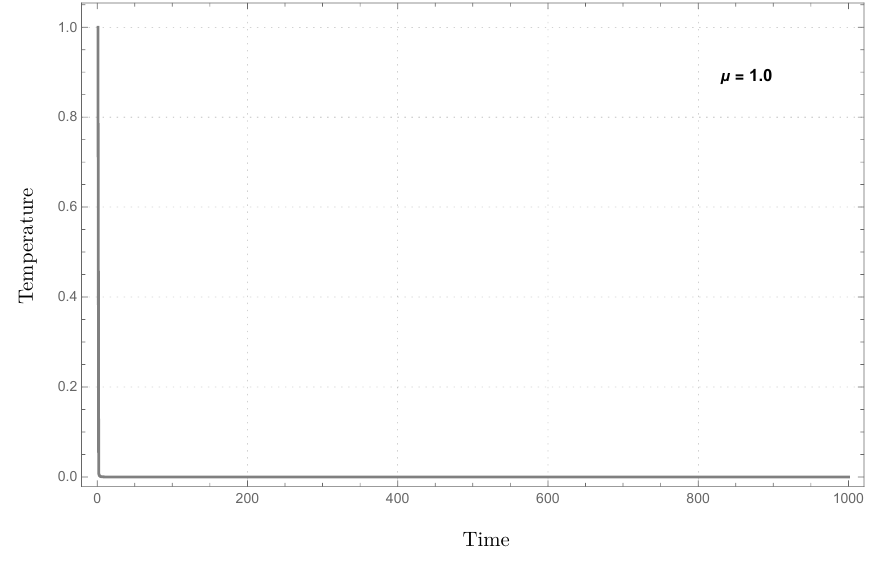}
\end{minipage}
\begin{minipage}[h]{1\linewidth}
\includegraphics[width=\linewidth]{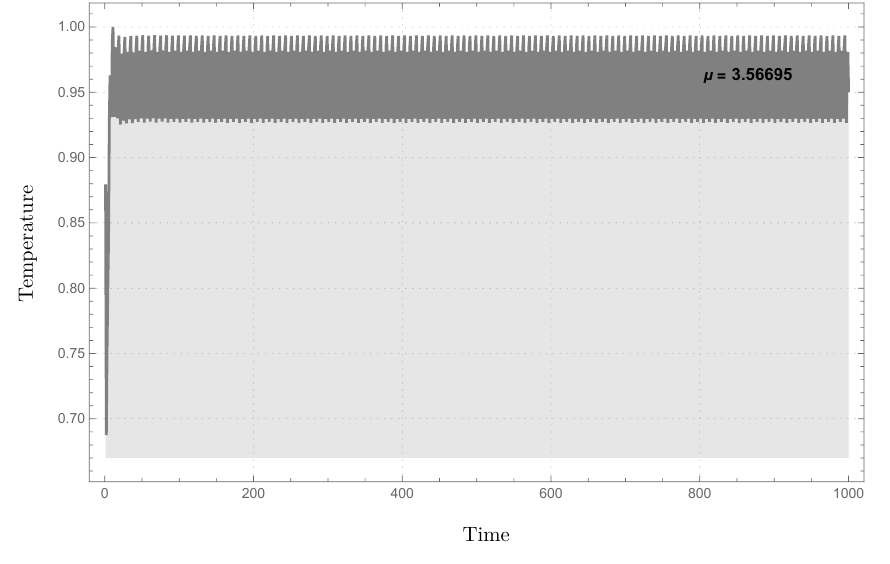}
\end{minipage}
\begin{minipage}[h]{1\linewidth}
\includegraphics[width=\linewidth]{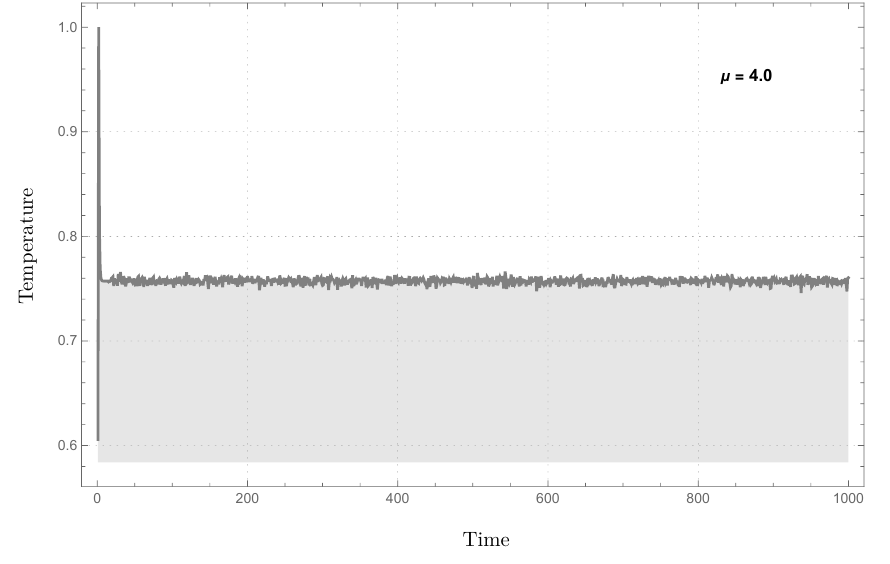}
\end{minipage}
\caption{Map temperature $T(\mu,N,M)$ versus time step of the logistic map for the parameter values $\mu=1$ (upper), $\mu=3.56695$ (center) and $\mu=4$ (bottom) provided with $M=10^5$ initial conditions.}
\end{figure}

\subsection{Fisher information and CR complexity as a function of temperature}
Using the Liouville canonical distribution $\rho(q,p)$ expressed as a product of the position and momentum distributions, in \cite{Plastino} it has been shown that the Fisher information of  $\rho(q,p)$ is a decreasing monotonic function of temperature. Inspired by this study and considering that the MP reaches an equilibrium value as if they were $M$ particles in contact with a thermal reservoir, now we investigate the FI and the CR complexity as functions of the temperature. 
From the upper panel of Fig. 8 we see that the Fisher information behaves predominantly within a concentrated strain of values when the temperature is near to the maximum $T=1$. We observe a peak of the FI near to the zero temperature, corresponding to the regular dynamics. Also, greater values of the FI are matched with more regular regimes. 
All this seems to be in agreement with the fact that nonzero map temperatures must be associated to a dynamics with some level of chaos, in order to be consistent with the idea of individual trajectories wandering over the all state space, thus corresponding to lower values of the Fisher information. Contrarily, for regular regimes $0<\mu\leq 3$ the orbits converge to a limit $x_\infty$, thus implying  
a null average velocity $x_{N+1}^{(j)}-x_N^{(j)}\sim 0$ in the limit $N\rightarrow \infty$ of large iterations for all initial condition $j=1,\ldots,M$ and then resulting in a zero temperature $T(\mu,N,M)$ and in a greater value of the Fisher information.
From the lower panel in Fig. 8, we see consistently that lower values of the CR complexity are associated to the regimes with a simple dynamic like the regular and fully chaotic ones. The peak in the region $0.8\leq T\leq 1$ corresponds to the onset of chaos zone, thus expressing the effectiveness of the CR complexity in detecting complex behavior. 

\begin{figure}
\begin{minipage}[h]{1\linewidth}
\includegraphics[width=\linewidth ]{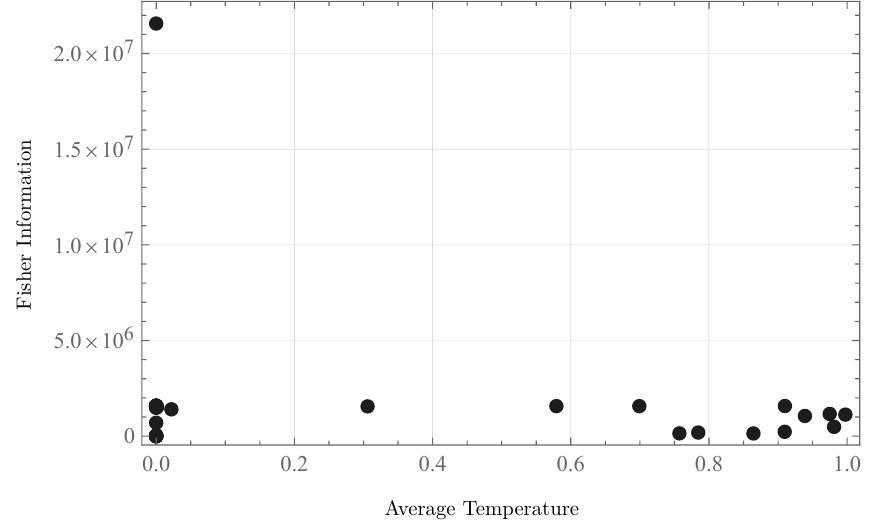}
\end{minipage}
\begin{minipage}[h]{1\linewidth}
\includegraphics[width=\linewidth]{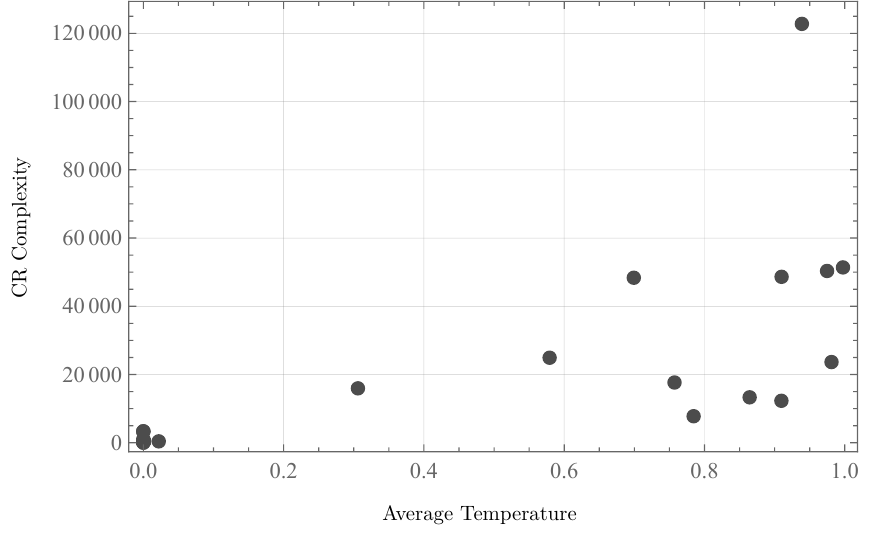}
\end{minipage}
\caption{Fisher information and Crámer-Rao complexity versus normalized average temperature \eqref{MP} of the logistic map after $N=10^3$ time steps and for $M=10^5$ initial conditions.}
\end{figure}

\section{Conclusions}
We have presented a study of the logistic map from FI and CR complexity numerical calculations for the invariant density distribution and, by employing an analogy with the Equipartition Theorem, we have defined an associated map temperature. Our contributions
are enumerated as follows. 
\begin{itemize}
    \item[(a)] The Fisher information and the Crámer-Rao complexity of the invariant density allow to distinguish regimes and transition dynamics in terms of the parameter of the logistic map, as evidenced in Figs. 1 and 2. In particular, the FI is useful for detecting regular regimes expressed by peaks due to its sensitivity, while the CR complexity characterizes regular, chaotic and complex regimes with a presence of peaks in transitional regions like the onset of chaos, Pameau-Maneville scenario. In addition, the graph FI versus CR complexity can be employed for detecting a correlation between them in order to identify regions of complex behavior (Fig. 3).        
    \item[(b)] The time evolution of the FI and of the CR complexity of the invariant density characterize several features of the dynamics for representative values of the parameter of the map, as we can see from Figs. 4 and 5, and provided with a small number of iterations. For the regular regime $\mu=1$, the presence of intermittencies with a decreasing and an abrupt grow represent a violation of the Frieden's informational intepretation of the Second Law \eqref{FI-arrow}. The CR complexity for $\mu=1$ expresses a decreasing monotonic behavior, thus suggestively indicating that when the Frieden's condition $dI/dt\leq 0$ is dissatisfied, the CR complexity still preserves the Frieden's interpretation with $dC/dt\leq 0$ and in agreement with the fact that regular regimes must correspond to a vanishingly small complexity. When $\mu=3.56695$ and $\mu=4$, fluctuations and oscillations are manifested both for the FI and for the CR complexity, indicating that these statistical quantifiers do not fully stabilize during chaotic behavior.     
    \item[(c)] The map temperature MP proposed in \eqref{MP}, arising from an analogy between the velocities of the particles of a gas and the mean velocity between two consecutive positions for several initial conditions, is a good quantifier of the dynamics, as we can see from Fig. 6. With low number of iterations $N=10^3$, we observe that for all the regular regime $0<\mu \leq 3$ the map temperature is zero, consistent with the convergent orbits for all initial condition. In the region $3<\mu\leq 4$ the MP is nonzero with a non monotonic from $\mu=3$ until $\mu=4$ with the presence of variations in the complex regime (Pameau-Maneville scenario) $3.56995\leq \mu \leq 3.82843$ and then decreasing up to $\mu=4$. 
    The MP in function of the time has evidenced an asymptotic stabilization only for regimes with low complexity, i.e. for $\mu=1$ (regular) and $\mu=4$ (fully chaotic) while for the onset of chaos $\mu=3.56695$ 
    the fluctuations impede the reaching of an equilibrium temperature.
    \item[(d)] From the MP \eqref{MP}, we have analyzed the FI and the CR complexity as functions of the temperature in light of \cite{Plastino},
    illustrated in Fig. 8. By means of an analogy between particles in the canonical ensemble at an equilibrium temperature and the MP originated by the dynamics of the initial conditions, we have found that for the invariant density of the logistic map the condition $dI/dT\leq 0$ is not guaranteed, mainly due the complex dynamics of the transition region (onset of chaos) corresponding to the agglomerated points in $0.8\leq T\leq 1$.  Maximum values of the FI and the CR complexity detect temperatures where the behavior is regular and complex. 
\end{itemize}
The present study of the logistic map by means of the FI and CR complexity of the invariant density, along with our definition of map temperature, has been shown to be consistent in characterizing regular, fully chaotic and complex regimes, with an additional advantage   
of requiring smaller number of iterations and of initial conditions than other ones employed in the literature.
A second benefit of our study is the theory of Frobenius Perron and Koopman operators of the densities connecting the numerical calculations of the invariant density.
Finally, a third advantage is that the MP only requires the calculation of the mean kinetic energy 
for a relatively small number of iterations in order to display dynamical features.
The main results of our study are illustrated in a summarized way in Table 1. 
In future researches we hope to apply the formalism presented in other chaotic maps \cite{Guinness,Boyland,Glass}.  

\begin{table}[]
    \centering
    \begin{tabular}{|c|c|c|c|}
    \hline
    & parameter $\mu$ & $T(\mu,N,M)$ & time $t$ \\
    \hline
        \textbf{Fisher} & localized  & delimited & Frieden's\\
         \textbf{information} & at $\mu=1$ & in a strip & condition \eqref{FI-arrow}\\
        $I[p]$ \eqref{Fisher-information} & & and & asymptotically  \\
        & & concentrated & with \\
        & & around the & $dI/dt\sim 0$ and \\
        & & maximum &  fluctuations,\\
        & & value $T=1$& but violated \\
        & & & for $\mu=1$ \\
        \hline
        \textbf{Crámer-Rao}  & sensitive & neither & asymptotically\\
        \textbf{complexity} & variations & decreasing & constant\\
        $C[p]$ \eqref{CR} & in $3\leq \mu \leq 4$ & nor & with \\
        & & increasing & $dC/dt\sim 0$\\
        \hline
        \textbf{map} & $\sim 0$ until & & zero for\\
        \textbf{temperature} & $\mu=3$, where & & $\mu=1$ and \\
        $T(\mu,N,M)$ & it starts to & & nonzero for\\
       \eqref{MP} & grow, variations & & $\mu=4$ with\\
        & in $3\leq \mu \leq 4$ & & fluctuations\\
        &&& for $\mu=3.56695$ \\
        \hline
    \end{tabular}
    \caption{A summary of the main characteristics of the Fisher information, CR complexity and map temperature of the logistic map as functions of the  parameter $\mu$, the map temperature $T(\mu,N,M)$ and the time step $t$.}
    \label{tab:placeholder}
\end{table}


\section*{Acknowledgments}
Samuel S. Santos, Guilherme V. B. Junior,  Ignacio S. Gomez and Ronaldo Thibes acknowledge support from the Department of Exact
and Natural Sciences of the State University of Southwest
Bahia (UESB), Itapetinga, Bahia, Brazil. Ignacio S. Gomez acknowledges support received from the Conselho Nacional de Desenvolvimento Científico e Tecnológico (CNPq), Grant Number
316131/2023-7. Guilherme Vieira Brito Junior acknowledges support received from FAPESB as a scientific initiation scholarship.

\end{document}